\documentclass{aa}  
\usepackage{natbib}
\usepackage{graphicx}
\usepackage{txfonts}
\begin{document}
   \title{Relationship between photospheric currents and \\ coronal magnetic
   helicity for force-free bipolar fields}

   \author{S. R\'egnier
          }

   \offprints{S. R\'egnier}

   \institute{School of Mathematics and Statistics, University of St Andrews, St Andrews,
   Fife, KY16 9SS, UK\\
              \email{stephane@mcs.st-andrews.ac.uk}
             }

   \date{Received ; accepted }

 
  \abstract
   {}
   {The origin and evolution of the magnetic helicity in the solar corona are not well
   understood. For instance, the magnetic helicity of an active region is often about
   $10^{42}$ Mx$^2$ ($10^{26}$ Wb$^{2}$), but the observed processes whereby it is thought
   to be injected into the corona do not yet provide an accurate estimate of the resulting
   magnetic helicity budget or time evolution. The variation in magnetic helicity is
   important for understanding the physics of flares, coronal mass ejections, and their
   associated magnetic clouds. To shed light on this topic, we investigate here the
   changes in magnetic helicity due to electric currents in the corona for a single
   twisted flux tube that may model characteristic coronal structures such as active
   region filaments, sigmoids, or coronal loops.}
   {For a bipolar photospheric magnetic field and several distributions of
   current, we extrapolated the coronal field as a nonlinear force-free field. We
   then computed the relative magnetic helicity, as well as the self and mutual
   helicities.}
   {Starting from a magnetic configuration with a moderate amount of current,
   the amount of magnetic helicity can increase by 2 orders of magnitude when
   the maximum current strength is increased by a factor of 2. The high
   sensitivity of magnetic helicity to the current density can partially explain
   discrepancies between measured values on the photosphere, in the corona, and
   in magnetic clouds. Our conclusion is that the magnetic helicity strongly
   depends on both the strength of the current density and also on its
   distribution.}
   {Only improved measurements of current density at the photospheric level will
   advance our knowledge of the magnetic helicity content in the solar
   atmosphere.}


   \maketitle

\section{Introduction}


Magnetic helicity is an important quantity in the physics of eruptive events
occurring in the solar corona. Active region filaments and sigmoids have been
observed and modelled as twisted flux bundles \citep{rus96, aul98a, can99,
aul99, reg02, gib02, reg04, tor03}. Magnetic helicity is also thought to play an
important role in the formation of filaments \citep[e. g.,][]{mac97, mac05} and
possibly in coronal heating \citep[e. g.,][]{hey84, pri99}. The instability of
coronal structures is related to the amount of helicity stored in the magnetic
field and the possible transfer of twist to writhe or self helicity from mutual
helicity. 

Nevertheless, only proxies of the magnetic helicity have been used to
estimate the twist of magnetic structures. In \cite{lea04}, the linear
force-free parameter $\alpha$ was derived as a proxy for the twist of coronal
structures observed in X-rays based on the thin flux tube approximation. The
twist values of the coronal structures and of the associated magnetic clouds are
inconsistent, suggesting that the propagation of coronal mass ejections (CMEs)
in the corona involves a transfer between the mutual helicity of the overlying
field and the self helicity of the flux rope. Based on a linear force-free
assumption, \cite{dem02a} have shown that the helicity of magnetic clouds is
comparable to the end-to-end helicity of the twisted bundle formed in the
associated active region. This suggests that the helicity is more likely to be
injected through the photosphere by flux emergence or localized magnetic field
motions \citep[see also][]{cha01a, kus02, wel03, mag03, lon07}. The helicity
injection processes play a key role in the long-term evolution of solar magnetic
field \citep[e. g., ][]{yea08}. The helicity of magnetic clouds, however, is
often found to be one or two orders of magnitude greater than the estimated
coronal helicity. In this Letter, we model coronal structures by a single
twisted flux tube with several distributions of current density and study the
variations in the helicity content due to an increase in current density.  

In reconstructed magnetic fields, the magnetic energy increases when the current
density is increased; however, we can only inject a finite  amount of current into
a finite domain of computation. An upper limit of the amount of free
magnetic energy is given by the Aly-Sturrock (AS) limit \citep{aly84, stu91}
stating that, for a magnetic field strength decaying fast enough at infinity in
the half space above the photosphere, the magnetic energy of the open magnetic
field is the least upper bound of the magnetic energy of a force-free field.
Thus, the magnetic helicity that can be injected into a magnetic configuration is
also bounded. The magnetic energy of the open field configuration is about twice
the magnetic energy of the potential field \cite[e.g.,][]{ama00}. Both the
potential field energy and open field energy depend of the total unsigned flux
magnetic field through the surface. In a finite domain of computation as is
the case for magnetic field extrapolations, the AS limit is used to check the
validity of a magnetic configuration, therefore the extrapolated configuration is
considered valid when the magnetic energy of the force-free field is lower than
the magnetic energy of the open field. This condition is true when closed
boundary conditions are used to derive the nonlinear force-free field. 

\section{Magnetic field and current distributions}
\label{sec:obs_mod}

The first step in order to derive the magnetic helicity content in the corona is
to compute the 3D magnetic field in a finite volume. We assume that the coronal
magnetic field is described well by a nonlinear force-free field satisfying the
following equations:
$\vec \nabla \wedge \vec B = \alpha \vec B$,
where $\alpha$ is the force-free function depending on the position, and
$\vec B \cdot \nabla \alpha = 0$,
which implies that $\alpha$ is a constant along a given field line. The
magnetic field also has to satisfy the solenoidal condition ($\vec \nabla \cdot
\vec B = \vec 0$). We solve this problem in accordance with the method
developed by \cite{gra58}. Following \cite{sak81}, the boundary conditions for
solving this set of equations as a mathematically 
well-posed boundary problem are the vertical
component of the magnetic field everywhere on the surface $\delta \Omega$, 
and the distribution
of $\alpha$ in one chosen polarity on $\delta \Omega^{\pm}$. We use the 
numerical scheme developed by
\citeauthor{ama97} (\citeyear{ama97, ama99b}), which  was successfully applied
to solar active regions by \citeauthor{reg02} (\citeyear{reg02, reg04, reg06})
by qualitatively comparing the computed field lines to multi-wavelength
observations.
It is important to note that we use closed boundary conditions on the sides of
the computational box, different from the bottom boundary, which are compatible
with the boundary conditions used to derive the helicity integrals (see
Section~\ref{sec:model}).    

\begin{figure}
\centering
\begin{minipage}[c]{.49\linewidth}
\includegraphics[width=1.\linewidth]{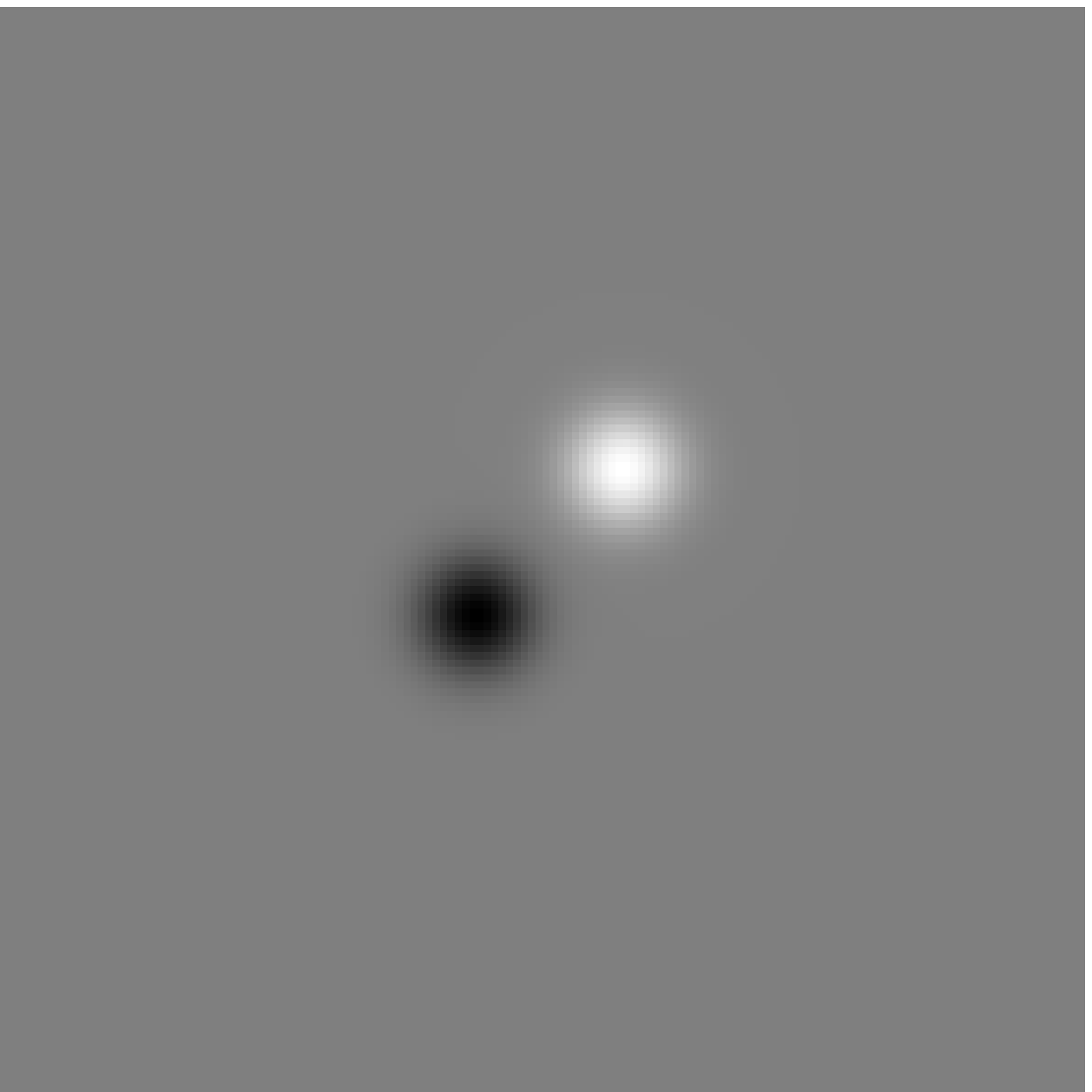}
\end{minipage}
\begin{minipage}[c]{.49\linewidth}
\includegraphics[width=.49\linewidth]{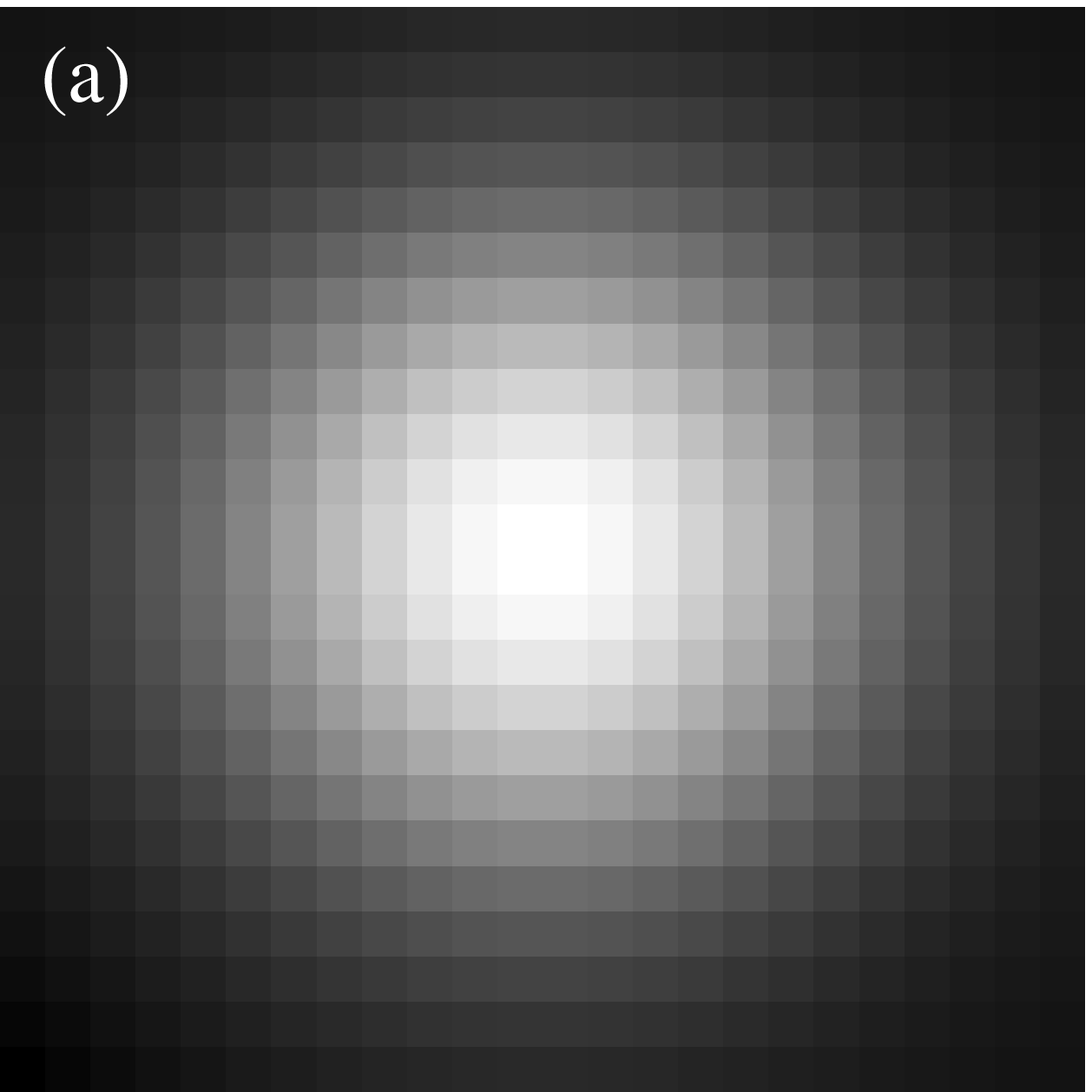}
\includegraphics[width=.49\linewidth]{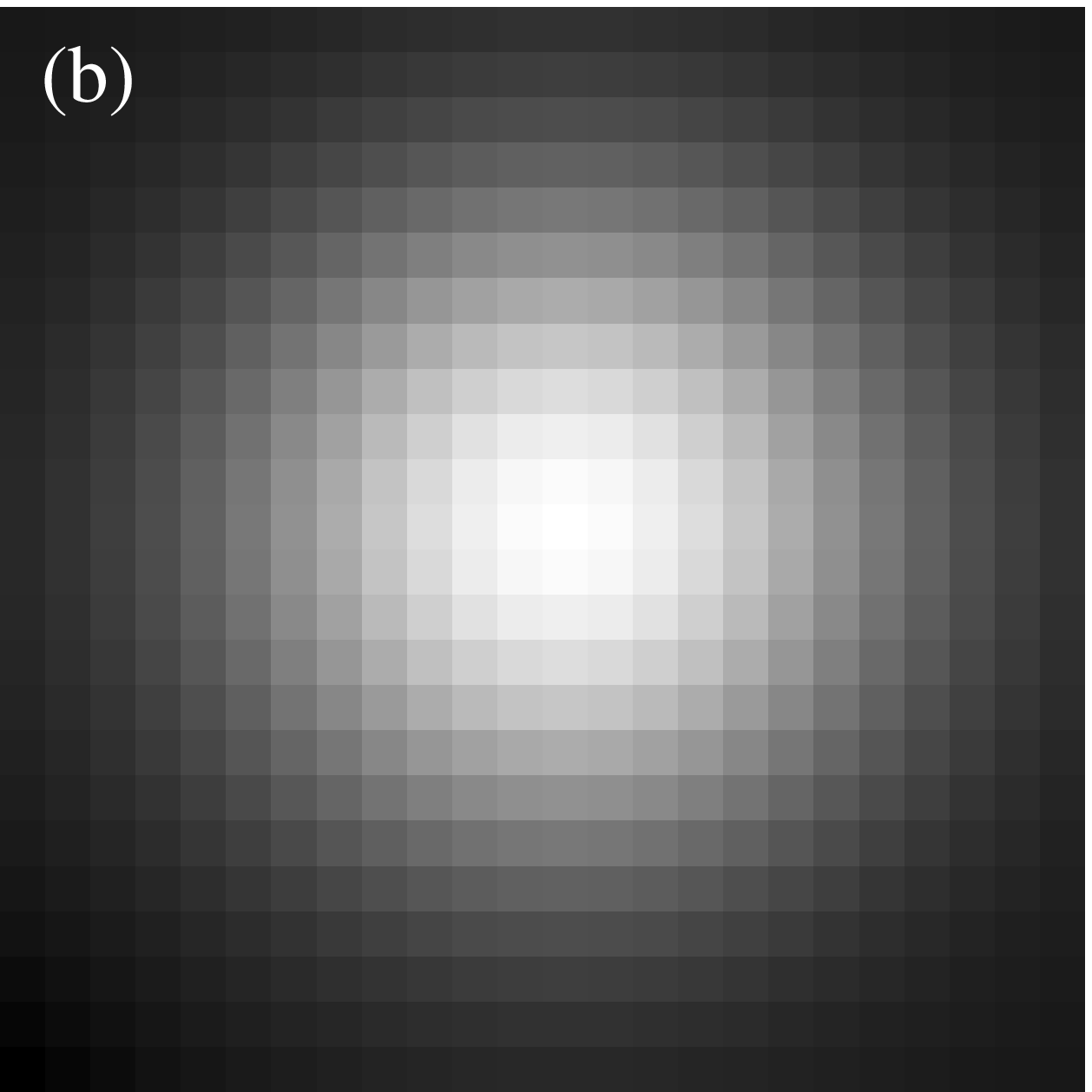}
\includegraphics[width=.49\linewidth]{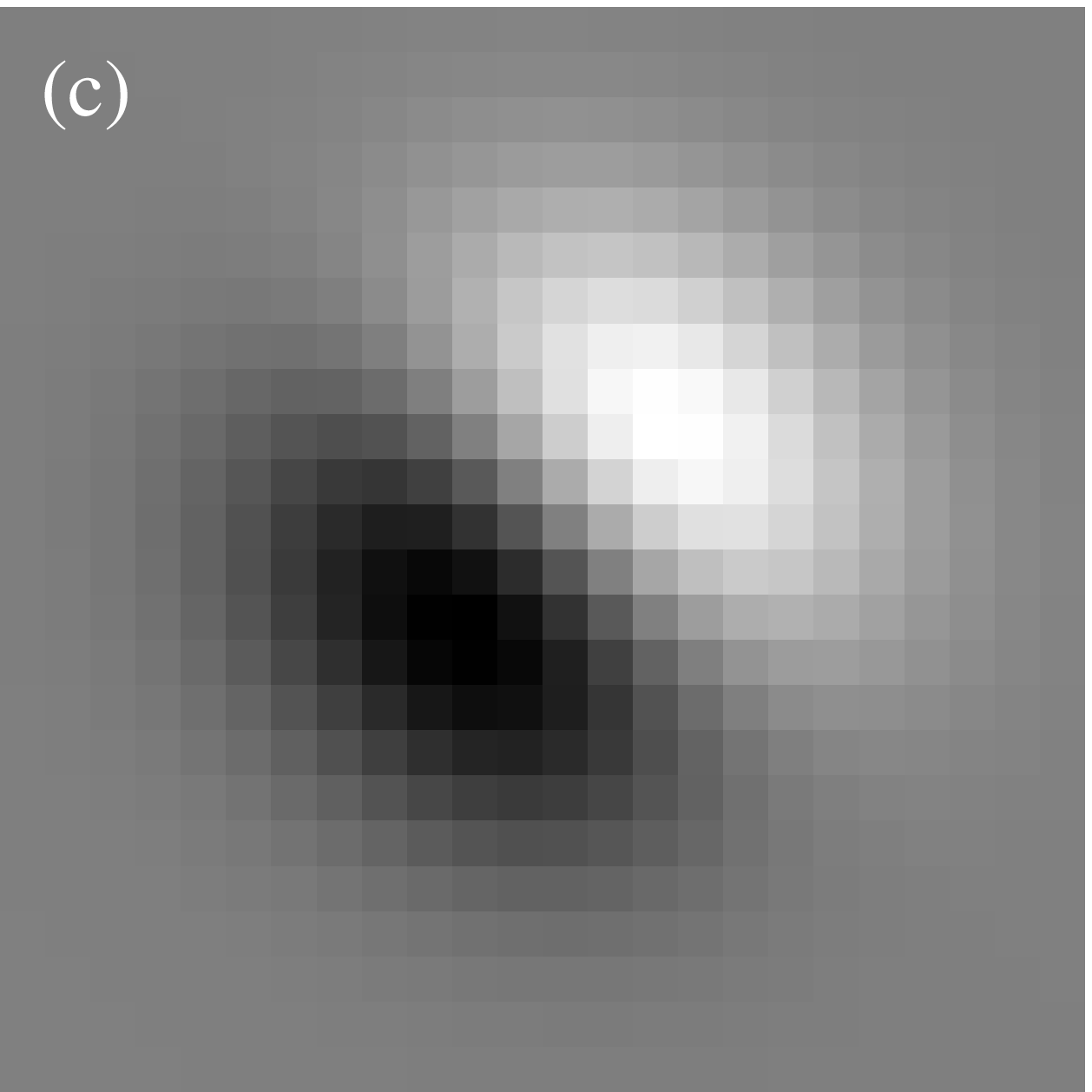}
\includegraphics[width=.49\linewidth]{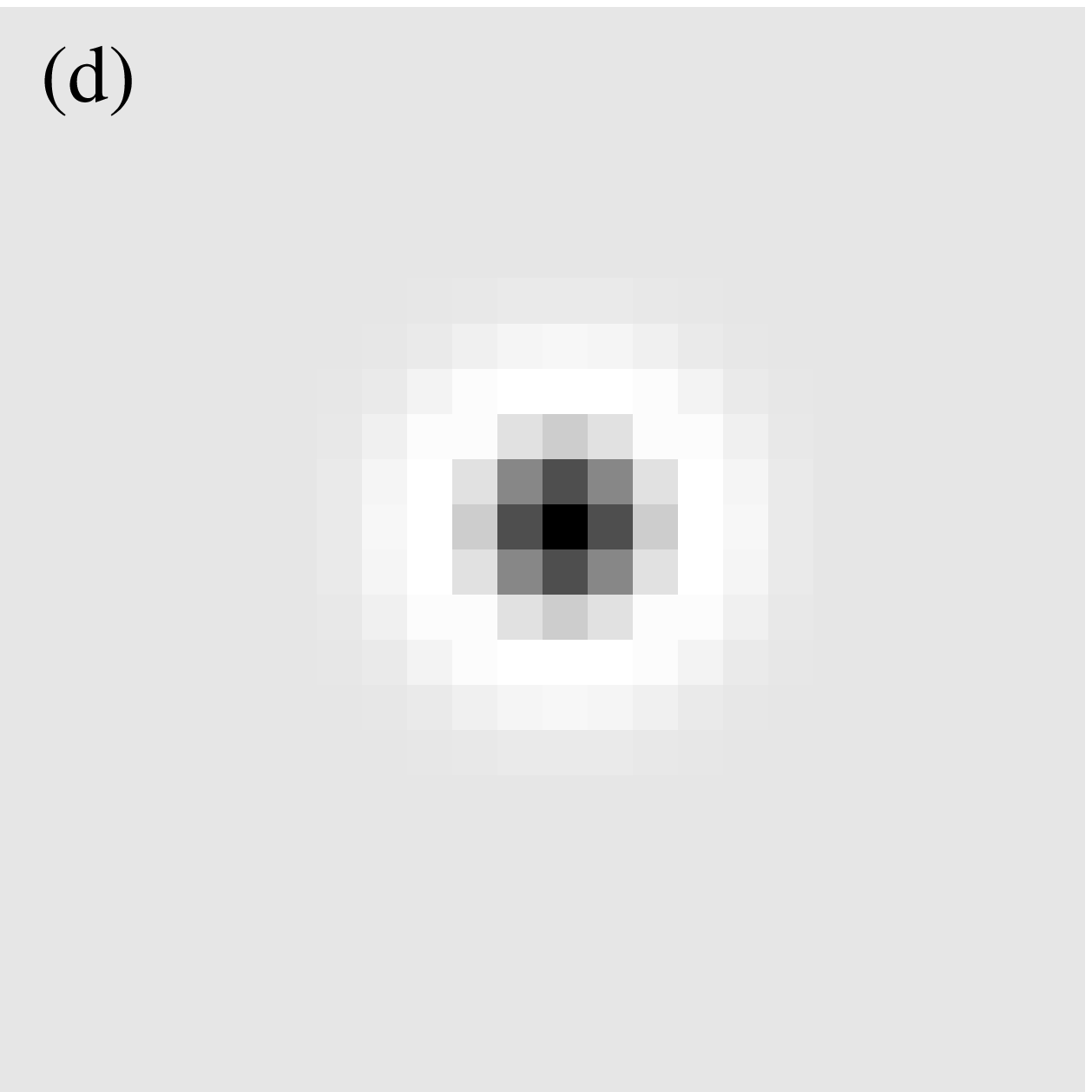}
\end{minipage}
\caption{Left: distribution of the vertical component of the magnetic field on
the bottom boundary. The polarities are defined as Gaussian distributions with
the same maximum strength in absolute value and the same FWHM. The field-of-view
is 150$\times$150 Mm$^{2}$. The negative (positive) polarity is black
(white). Right: (a) vertical magnetic field component in the positive polarity;
(b) Gaussian distribution of the vertical current corresponding to a constant
$\alpha$ field; (c) divided distribution of current; (d) ring distribution of
current.}
\label{fig:bz_dip}
\end{figure}

The second step is to define the appropriate boundary conditions on the bottom boundary:
the vertical magnetic field on the surface and the distribution of the force-free function
$\alpha$ in one chosen polarity. As depicted in Fig.~\ref{fig:bz_dip} left, the vertical
magnetic field on the bottom boundary is a bipolar field embedded in a large field-of-view
in order to minimise the influence of the boundaries on the magnetic field configurations.
The vertical magnetic field component $B_z$ is described by a Gaussian distribution (see
Fig.~\ref{fig:bz_dip}a) with a maximum field strength of 2000 G and a full-width at
half-maximum (FWHM) of  about 15 Mm. The field-of-view is 150$\times$150 Mm$^2$ with a
spatial resolution of 1 Mm. The peak-to-peak separation of the polarities is about 30 Mm. 
	
The vertical current distributions are defined in the positive polarity as
follows:
\begin{itemize}
\item[-]{{\em Constant}: $\alpha$ is a constant (see Fig.~\ref{fig:bz_dip}b)
corresponding to a linear force-free field. The current distribution is then a
Gaussian distribution with the same FWHM as $B_z$ and a maximum current density 
strength $J_{z0}$.} 

\item[-]{{\em Divided}: the distribution is defined from a Hermite polynomial
of 1st order dividing the polarity into a negative and a positive part (see
Fig.~\ref{fig:bz_dip}c):
\begin{equation}
J_{z}(r) = 2~J_{z0}~r~\exp{\left(-\frac{r^2}{\sigma^2}\right)},
\end{equation}
where $r$ is the distance from the centre of the polarity. The current density 
flux through the positive polarity at the bottom boundary is then balanced. A
free parameter of the current distribution is the angle $\theta$ between the
polarity inversion line (PIL) and the current inversion line (CIL) in the
positive polarity. We have chosen $\theta = 0$ (with the negative currents
towards the PIL) such that the magnetic energy is maximised.
}

\item[-]{{\em Ring}: the vertical current density is defined as a Hermite
polynomial of 2nd order in 1D:
\begin{equation}
J_{z}(r) = 2~J_{z0}~(r^2 - C_0)~\exp{\left(- \frac{r^2}{\sigma^2}\right)}
\end{equation}
where $r$ is the distance from the centre of the polarity. A typical ring
distribution is plotted in Fig.~\ref{fig:bz_dip}d with negative currents in the
central region surrounded by positive currents. The constant $C_0$ is such that
the vertical current density in the positive polarity is balanced.}
\end{itemize}

\section{Helicity measurements}
\label{sec:model}

	\subsection{Magnetic helicities}

The magnetic helicity describes the complexity of the field in terms of its
topology, connectivity or braiding, and it is a measure of both the twist
of field lines around the flux bundle axis and writhe of the axis itself
\citep{ber99}.
The magnetic helicity is a conserved quantity in ideal MHD for a volume bounded
by a surface on which the normal field component is fixed. However, the magnetic
helicity is not conserved whilst modelling the solar corona above the photosphere
since helicity may be injected from below the photosphere into the corona
\citep[e.g.,][]{reg06} or it may be expelled in magnetic clouds during CMEs into
the interplanetary  medium.

The magnetic helicity is defined as follows:
\begin{equation}
H_m(\vec B) = \int_{\Omega}~\vec A \cdot \vec B~d\Omega
\end{equation}
for a magnetic field $\vec B$ and its vector potential $\vec A$ in a volume
$\Omega$. The vector potential is not defined uniquely but depends on a
gauge. Here we use a gauge-free expression of the
magnetic helicity due to \cite{ber84} and called the relative magnetic helicity
\begin{equation}
\Delta H_m(\vec B, \vec B_{pot}) = \int_{\Omega}~( \vec A - \vec A_{pot}) \cdot
( \vec B + \vec B_{pot})~d\Omega,
\end{equation}
where $\vec B$ and $\vec A$ describe the magnetic field of the configuration,
and, $\vec B_{pot}$ and $\vec A_{pot}$ describe a reference field taken to be
the potential field. We use the same boundary conditions as for the universal
helicity formula derived by \citet{hor06}.  

Following \cite{ber99}, we define the self and mutual helicities as
\begin{equation}
H_{self}(\vec B_{cl}) = \int_{\Omega}~\vec A_{cl} \cdot \vec B_{cl}~d\Omega
\end{equation}
and
\begin{equation}
H_{mut}(\vec B_{pot}, \vec B_{cl}) = 2~\int_{\Omega}~\vec A_{pot} \cdot \vec
B_{cl}~d\Omega
\end{equation}
when the field $\vec B$ can be decomposed into two fields, the reference field
$\vec B_{pot}$ and the closed field $\vec B_{cl}$. The boundary conditions are
explained in \cite{reg05} and they are the same as used to compute the nonlinear
force-free field in this experiment. 

We note that those definitions of self and mutual helicities are different from
the recent definitions given by \citet{lon08a}. In the solar context, the self
and mutual helicities as derived by \cite{ber99} have been defined in
\cite{reg05} from measurements based on simple configurations and observed
active regions. The self helicity is a measure of the twist and writhe of flux
bundles confined in the coronal volume. The mutual helicity characterises the
crossing of field lines and the large-scale twist. 

	\subsection{Helicities vs. current}

\begin{figure}
\centering
\includegraphics[width=.8\linewidth]{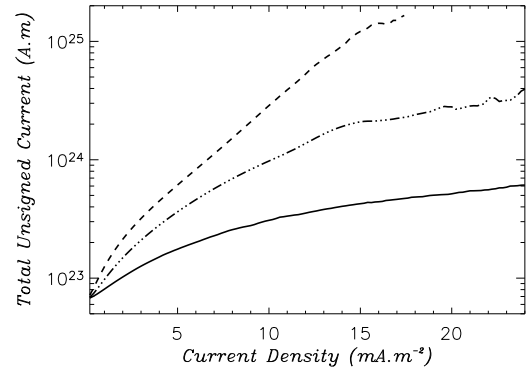}
\caption{Total unsigned current (A$\cdot$m) inside the computational volume as 
a function of the current density $J_{z0}$ for a constant distribution (dashed
line), a divided distribution (dot-dashed line), and a ring distribution (solid
line).}
\label{fig:tot_j}
\end{figure}

\begin{figure*}
\centering
\includegraphics[width=.32\linewidth]{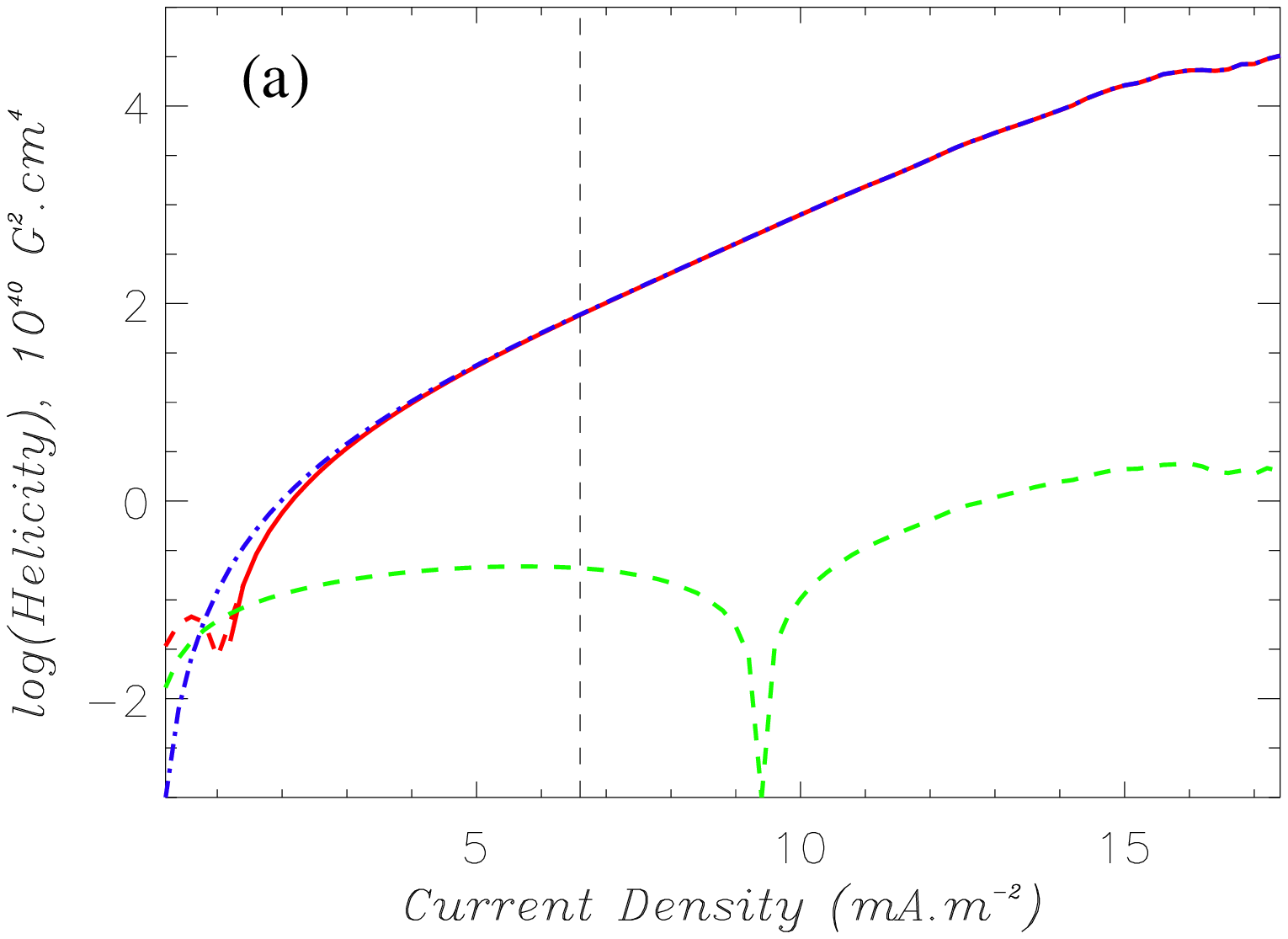}
\includegraphics[width=.32\linewidth]{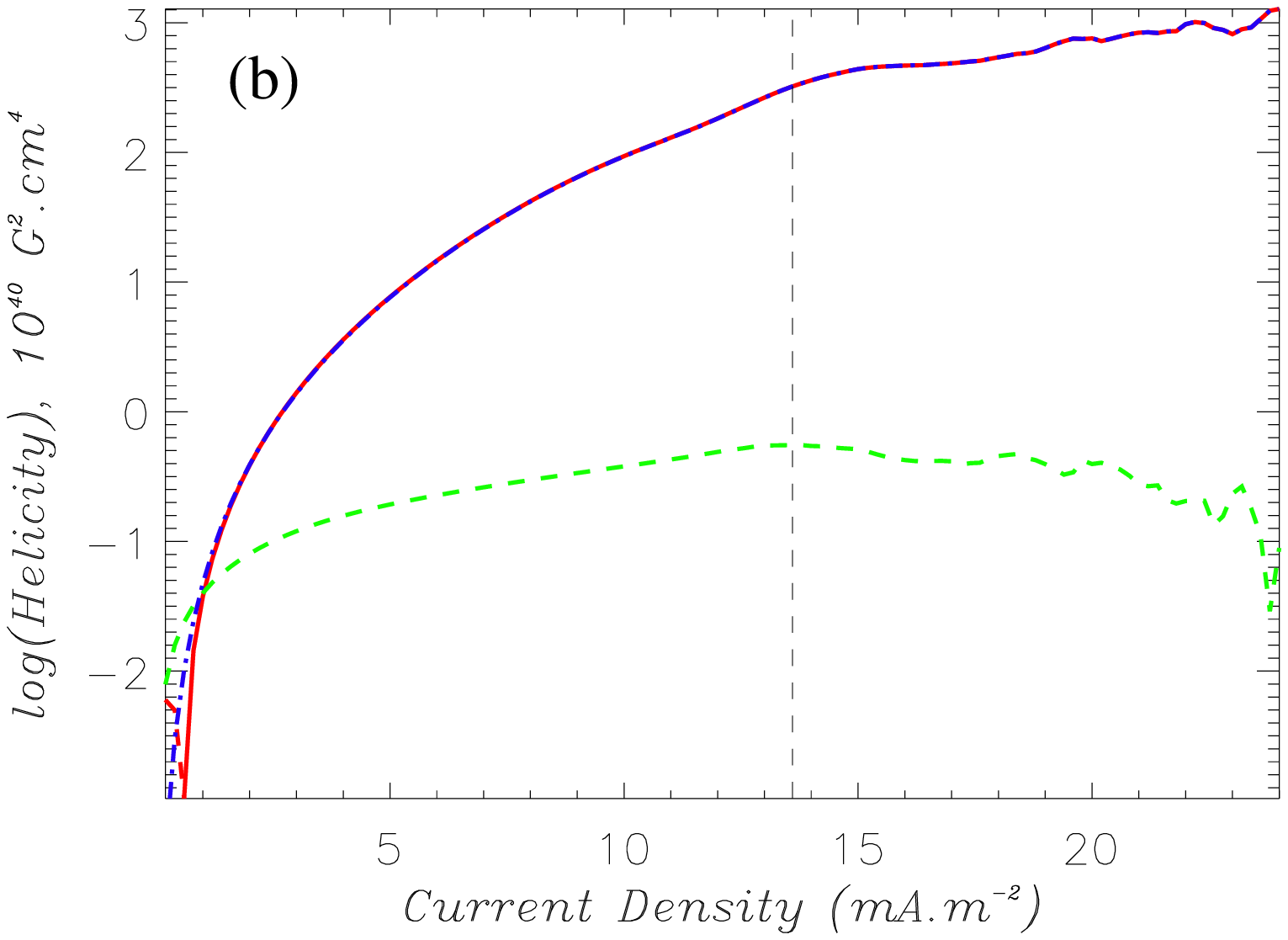}
\includegraphics[width=.32\linewidth]{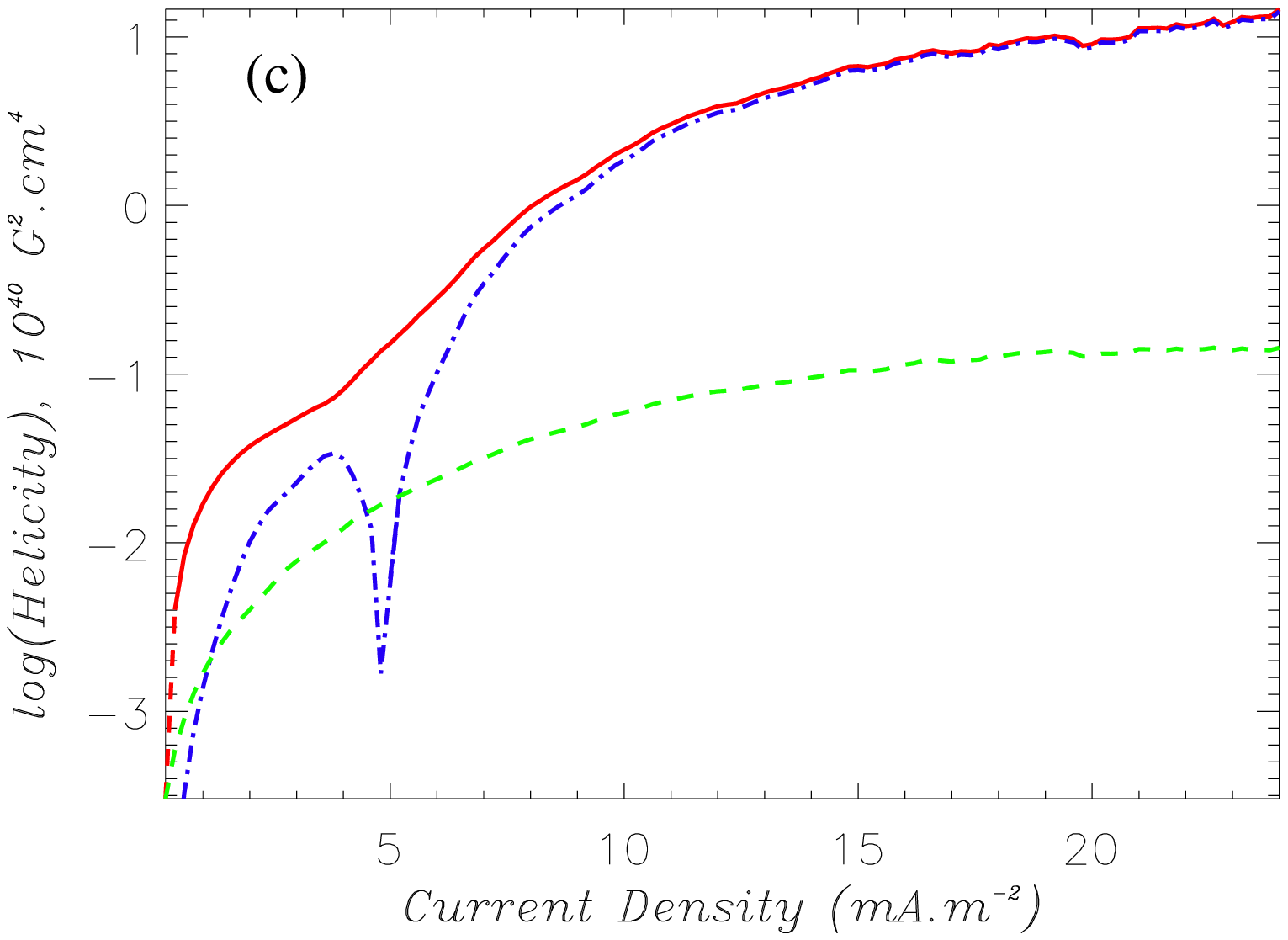}
\caption{Relative magnetic helicity ({\em red} solid line), self helicity ({\em
blue} dot-dashed line), and mutual helicity ({\em green} dashed line) as a
function of the maximum vertical current density $J_{z0}$ from 0 to 24
mA$\cdot$m$^{-2}$ for (a) a constant distribution, (b) a divided distribution,
(c) a ring distribution.  The helicities are expressed in units of 10$^{40}$
G$^2\cdot$cm$^{4}$. The straight dashed lines indicate the Aly-Sturrock limit.}
\label{fig:jz_plot}
\end{figure*}

We now compute the nonlinear force-free field in the corona for the three
different distributions of current, and we study the changes in the magnetic
configurations caused by an increase in the maximum vertical current strength 
$J_{z0}$. In a forthcoming paper, we will extensively study the changes in the
geometry of field lines, the magnetic connectivity, and the magnetic energy
budget. In this Letter, we focus our study on the changes in the magnetic
helicity content of the bipolar field for the three different current
distributions described in Section~\ref{sec:obs_mod}. The maximum current strength
ranges from 0 to 24 mA$\cdot$m$^{-2}$. We first
plot the total unsigned current inside $\Omega$ as a
function of $J_{z0}$ in Fig.~\ref{fig:tot_j} for the three current distributions. In
Fig.~\ref{fig:jz_plot}, the self and mutual helicities are the blue and green
curves, respectively, whilst the total relative magnetic helicity is the red
curve. We also indicate the AS limit when the magnetic
energy of the nonlinear force-free field is 1.7 times the magnetic energy of the
potential field, corresponding to the magnetic energy computed for the open
field. This upper limit  gives $J_{z0} = 6.6$ mA$\cdot$m$^{-2}$ for the
constant distribution, 13.6 mA$\cdot$m$^{-2}$ for the divided distribution. For
this range of $J_{z0}$ values, there is no upper limit for the ring distribution,
which indeed corresponds to a twisted flux bundle confined by return currents.

The magnetic helicities have a positive sign, except for the relative magnetic 
helicity of the ring distribution below $J_{z0} = 5$ mA$\cdot$m$^{-2}$ (see
Fig.~\ref{fig:jz_plot}c). From these computations, the mutual helicity values
are most sensitive to the existence of an upper bound for the magnetic energy.
For the constant and divided distributions (see Fig.~\ref{fig:jz_plot}a, b), the
behaviour of the mutual helicity is strongly modified above the AS limit,
whilst we get a smooth curve of mutual helicity for the ring distribution.

For all of the imposed current distributions, the magnetic helicity of the
bipolar field is dominated by the self helicity, and thus the configurations are twisted
flux tubes confined in a small domain of the computational box according to the
definition given in \cite{reg05}. The helicity values vary from 10$^{38}$ to
10$^{42}$ G$^2\cdot$cm$^4$ for the constant and divided distributions, from
10$^{37}$ to 10$^{41}$ G$^2\cdot$cm$^4$ for the ring distributions. The ring
distribution tends to reduce the amount of magnetic helicity stored in the
confined twisted flux tube. The relative and self helicities show an exponential
growth with increasing current density (linear trend in Fig.~\ref{fig:jz_plot}).
For moderate values of the current density, these helicities vary by more than 1
order of magnitude when the current density $J_{z0}$ is multiplied by a factor
of 2. For high values of  $J_{z0}$, a plateau exists for all helicities.
This suggests that instabilities such as kink instability can develop when the
maximum helicity is reached inside the twisted flux bundle. The increase in
magnetic helicity as a function of the total current is reduced compared to the
evolution with respect to $J_{z0}$, but it is still significant depending on the
distribution of $J_z$.

\section{Discussion and conclusions}
\label{sec:disc}

By modelling coronal magnetic structures by a bipolar field containing a single 
twisted flux tube, we have demonstrated that the magnetic helicity can vary by
more than an order of magnitude when the current density is increased by a
factor of 2. This shows that the departure from the potential field state has
important consequences on the amount of magnetic helicity that can be stored in
the corona, and therefore  on the helicity content expelled during CMEs. 

In terms of magnetic field modelling, we have shown that the behaviour of the
magnetic helicity in a nonlinear force-free field strongly depends on the chosen
distribution of current. If we inject a large amount of current density  within
the ring distribution, the magnetic helicity values are one order of magnitude
lower than for the divided distribution as the twisted flux tube remains
confined to  a small fraction of the coronal volume. This is caused by the
existence of return currents at the edges of the twisted flux bundle, whilst
for the constant and divided current distributions, the field lines can expand
towards the boundaries.

In terms of observation, the vertical current density is deduced from vector
magnetic field measurements in the photosphere (or in the chromosphere):
$J_{z}^{phot} = \frac{1}{\mu_0} \left( \frac{\partial B_{y}^{phot}}{\partial x} -
\frac{\partial B_{x}^{phot}}{\partial y} \right), 
$
The uncertainties on the vertical current density strongly depend on the noise
level of the transverse field components inverted from spectropolarimetric
observations. From the above conclusions, it is worth noticing that magnetic
helicity values in the corona obtained from photospheric observations have to be
considered with caution as a small error on the measurement of the field
components can dramatically change the value of the magnetic helicity. And
so to better understand the physics of twisted flux bundles in the
corona, it is important to improve the polarimetric resolution to increase the
signal-to-noise ratio on the transverse field components and to increase the
spatial resolution to resolve the current which is distributed on small scales
\cite[e.g.,][]{par96}. Even if the effects are weaker, these conclusions
also apply to the change in helicity as a function of the total unsigned current.

\begin{acknowledgements}
I thank E. R. Priest, D. Mackay, and G. Hornig for fruitful
discussions. I thank the UK STFC for financial support (STFC RG). The
computations were done using the XTRAPOL code developed by T. Amari (Ecole
Polytechnique, France). I also acknowledge the financial support by the
European Commission through the SOLAIRE network (MTRN-CT-2006-035484).
\end{acknowledgements}


\end{document}